\documentclass[12pt]{article}
\usepackage{amsmath}
\usepackage{graphicx,psfrag,epsf}
\usepackage{enumerate}
\usepackage{natbib}
\usepackage{url} 

\newcommand{\blind}{0}
\usepackage{arydshln}
\usepackage{multirow}
\usepackage{amsmath}
\usepackage{amssymb}
\usepackage{arydshln}
\usepackage{amsfonts}
\usepackage{algorithm}
\usepackage{algorithmic}
\usepackage{caption}
\usepackage{subfigure}
\usepackage{ntheorem}
\usepackage{enumerate}
\usepackage{verbatim}
\newtheorem{theorem}{Theorem}

\newtheorem{proposition}{Proposition}

\newtheorem{assumption}{Assumption}

\usepackage[colorlinks=true, allcolors=blue]{hyperref}
\usepackage{color}
\newcommand{\xbold}{\boldsymbol{x}}
\newcommand{\ybold}{\boldsymbol{y}}
\newcommand{\thetabold}{\boldsymbol{\theta}}
\newcommand{\eg}{\emph{e.g. }}

\begin{document}

\def\spacingset#1{\renewcommand{\baselinestretch}%
{#1}\small\normalsize} \spacingset{1}




\if0\blind
{
  \title{\bf  {Using early rejection Markov chain Monte Carlo and Gaussian processes to accelerate ABC methods}}
  \author{Xuefei Cao$^1$, 
    Shijia Wang$^{2*}$ 
    and
    Yongdao Zhou$^1$\thanks{
    Address correspondence to: Dr.\ Shijia Wang ({\tt wangshj1@shanghaitech.edu.cn}) and Dr.\ Yongdao Zhou ({\tt ydzhou@nankai.edu.cn}).}\\
    $^1$School of Statistics and Data Science, Nankai University, China\\
    $^2$Institute of Mathematical Sciences, ShanghaiTech University, China\\
    }
  \maketitle
} \fi

\if1\blind
{
  \bigskip
  \bigskip
  \bigskip
  \begin{center}
    {\LARGE\bf Using early rejection Markov chain Monte Carlo and Gaussian processes to accelerate ABC methods}
\end{center}
  \medskip
} \fi

\bigskip
\begin{abstract}
Approximate Bayesian computation (ABC) is a class of Bayesian inference algorithms that targets for problems with intractable or  {unavailable} likelihood function.  {It uses synthetic data drawn from the simulation model to approximate the posterior distribution.}
However, ABC is computationally intensive
for complex models in which simulating synthetic data is very expensive. In this article, we propose an early rejection Markov chain Monte Carlo (ejMCMC) sampler based on Gaussian processes to accelerate inference speed.
We early reject samples in the first stage of the kernel using a discrepancy model, in which the discrepancy between the simulated and observed data is modeled by Gaussian
process (GP). Hence, the synthetic data is generated only if the parameter space is worth
exploring. We demonstrate from theory, simulation experiments, and real data analysis that the new algorithm significantly improves inference efficiency compared to existing early-rejection MCMC algorithms. In addition, we employ our proposed method within an ABC sequential Monte Carlo (SMC) sampler.
In our numerical experiments, we use examples of ordinary differential equations, stochastic differential equations, and delay differential equations to demonstrate the effectiveness of the proposed algorithm.
We develop an R package that is available at \url{https://github.com/caofff/ejMCMC}.
\end{abstract}
\noindent%

\spacingset{1.45}

\section{Introduction}
\label{sec:intro}
In Bayesian statistics, we aim to infer the posterior distribution of the unknown parameter, which requires the evaluation of the likelihood function of data given a parameter value. However, in some complex statistical models (\eg stochastic differential equations), it is computationally very expensive or even not possible to derive an analytical formula of the likelihood function, but we are able to simulate synthetic data from the statistical model. Approximate Bayesian computation (ABC) is a class of Bayesian computation methods that avoids the computation of likelihood function and is suitable for this context. In ABC, we simulate synthetic data from the statistical model given a simulated parameter, and this parameter value will be retained if the simulated data sets are close enough to the observed data sets. 

Approximate Bayesian computation \citep{Pritchard1999Population,Beaumont2002ABC,sisson2018handbook} is originally introduced in the population genetics context. The simplest version of ABC is a rejection sampler. 
In practice,  {rejection sampling is inefficient, especially when there is a large difference between the prior distribution and the posterior distribution.}
\cite{Marjoram2003MCMC} introduced an ABC Markov chain Monte Carlo (MCMC) algorithm to approximate the posterior distribution. 
\cite{Wegmann2009MCMC} improved the performance of ABC-MCMC by relaxing the tolerance within MCMC while conducting subsampling and regression adjustment to the MCMC output.
\cite{Clarte2020Gibbs} explored a Gibbs type ABC algorithm that component-wisely targets the corresponding conditional posterior distributions.
Sequential Monte Carlo (SMC) methods \citep{doucet2001introduction, del2006sequential} have become good alternatives to MCMC for complex model inference. 
\cite{Sisson2007SMC, del2012adaptive} employed ABC-MCMC algorithms within the SMC framework of \cite{del2006sequential}.
\cite{Buchholz2019ABCQMC} improved ABC algorithm based on quasi-Monte Carlo sequences, and the resulting ABC estimates achieves a lower variance compared with Monte Carlo ABC. \cite{doi:10.1080/01621459.2022.2086132, Price_2017} implemented Bayesian synthetic likelihood to conduct  inference for complex models with intractable likelihood function. 
\cite{frazier2018asymptotic}  {studied} the posterior concentration property of the approximate Bayesian computation methods, and the asymptotic distribution of the posterior mean.

 {
The choice of summary statistics affects the inference efficiency and results of ABC. 
In general, we measure the discrepancy between two data sets by using the distance between their low-dimensional summary statistics. 
However, constructing low-dimensional summary statistics can be challenging due to lack of expert knowledge. \cite{sisson2011likelihood, marin2012approximate, blum2013comparative} provide a detailed summary of the selection of summary statistics in ABC. \cite{joyce2008approximately,nunes2010optimal}  select the best subset from a pre-selected set of candidate summary statistics based on various information criteria.
\cite{pudlo2016reliable} conduct selection of summary statistics via random forests. 
\cite{Fearnhead2012Summary} construct summary statistics in a semi-automatic fashion using regression.  
\cite{jiang2017learning, aakesson2021convolutional} learn summary statistics  automatically via neural networks.  
} 

 {For expensive ABC simulators, various methods have been proposed to increase the sample-efficiency of ABC. 
\cite{wilkinson2014accelerating} used Gaussian process (GP) to model uncertainty in likelihood to exclude regions with negligible posterior probabilities. \cite{Jarvenpaa2018GPABC,jarvenpaa2019Efficient,jarvenpaa2020batch} modelled the discrepancy between synthetic and observed data using GPs, and estimated the ABC posterior based on the fitted GP without further simulation (GP-ABC). 
\cite{Picchini2014SDE} proposed an early rejection ABC-MCMC algorithm for SDE parameter inference, in which the acceptance ratio of MH algorithm is divided into two parts. \cite{Everitt2021DASMC} used delayed acceptance (DA) MCMC \citep{christen2005markov} in ABC framework, and used cheap simulators to build a more efficient proposal distribution. }

In this article, we propose a new early rejection MCMC (ejMCMC) method to speed up inference for simulating expensive models, based on the  {seminal} work of \cite{Picchini2014SDE} and \cite{Jarvenpaa2018GPABC,jarvenpaa2019Efficient,jarvenpaa2020batch}. 
In the first step of our algorithm, we rule out the parameter space that is not worth exploring by running a Metropolis Hasting (MH) step with a predicted discrepancy value. Compared to alternative machine learning techniques, Gaussian processes furnish not just point predictions, but also encompass uncertainty estimates and exhibit efficacy with limited data sets. Hence, we choose a Gaussian process model to evaluate the discrepancy in this article.
 {Consequently, our proposed method can early reject samples in cases the algorithm of \cite{Picchini2014SDE} is inefficient (\eg  uniform priors, symmetric proposals).}
Then, we simulate data from the model only if it is highly possible to be accepted.
 {The posterior estimated by GP-ABC approaches \citep{Jarvenpaa2018GPABC,jarvenpaa2019Efficient,jarvenpaa2020batch} rely entirely on the fitted surrogate model. 
In practice, there exist cases that the discrepancy for parameters cannot be well modelled by GPs. For example,  if the shape of discrepancy  as a function of parameters is multi-modal, the Gaussian process model may not be able to accurately capture these peaks. The approximate Bayesian posterior obtained by the GP surrogate model could be inaccurate.  We show this phenomenon in Section \ref{subsec:toy}.
The proposed ejMCMC method effectively combines MCMC and GP discrepancy model, and hence can correct the inaccurate posterior estimated by GP-ABC approaches. 
The efficiency of DA-ABC approach \citep{Everitt2021DASMC} relies on a cheap
simulator used in proposing stage. 
It is challenging to build an effective but cheap simulator when there is a lack of expert knowledge. Our ejMCMC instead does not require as much prior knowledge.} 
Our numerical experiments indicate that the new method can achieve about $40\%$ acceleration compared with \cite{Picchini2014SDE}, and the estimated posterior is more accurate than \cite{Picchini2014SDE} with fixed computational budgets. We show that the ejMCMC based on GP discrepancy satisfies the detailed balance condition, and the posterior concentration property. The proposed algorithm is theoretically more efficient than the existing one. We also propose an efficient GP discrepancy function to balance the computational efficiency and ratio of false rejection samples. The ejMCMC can be designed as an efficient forward kernel within ABC-ASMC framework \citep{del2012adaptive}, and the resulting algorithm (ejASMC) is more efficient. 

The rest of article is organized as follows. In Section \ref{sec:method}, we show some background information of approximate Bayesian computation. In Section \ref{sec:ejm}, we introduce our proposed algorithms. In sections \ref{sec:sim} and \ref{sec:real}, we use a real data analysis and simulation studies to show the effectiveness of our method.
Section \ref{sec:conc} gives the conclusion, all proofs of the theoretical results are deferred to the Appendix.


\section{Approximate Bayesian computation methods}
\label{sec:method}
 {In this section, we provide a review of ABC algorithms relevant for this article. 
}
\subsection{Basics of Approximate Bayesian computation}

Let $\boldsymbol{\theta}$ be the parameter of interest, which belongs to a parameter space $\Theta \subset \mathbb{R}^p$, and let $\pi(\boldsymbol{\theta})$ be a prior density for $\boldsymbol{\theta}$. The likelihood function, denoted as $p(\cdot \mid \boldsymbol{\theta})$, is costly to compute, while it is possible to simulate data from the model with a given parameter $\boldsymbol{\theta}$. Consider a given observed data set $\ybold = (\ybold_{(1)}, \ybold_{(2)}, \ldots, \ybold_{(T)})^\top \in \mathcal{Y}^{T} \subset \mathbb{R}^{d \times T}$, where $d$ denotes the dimension of data and $T$ denotes the number of repeated measurements. The likelihood-free inference is achieved by augmenting the target posterior from  {$\pi(\thetabold\mid\ybold)\propto p(\ybold\mid\thetabold)\pi(\thetabold)$} to 
\begin{equation*}
\pi_\varepsilon(\thetabold,\xbold\mid\ybold)=\frac{\pi(\boldsymbol{\theta}) p(\xbold\mid\thetabold)K_\varepsilon(\Delta(\xbold,\ybold))}{\iint\pi(\boldsymbol{\theta}) p(\xbold\mid\thetabold)K_\varepsilon(\Delta(\xbold,\ybold))d\xbold d\thetabold},
\end{equation*}
where the auxiliary parameter $\xbold$ is a (simulated) data set from $p(\cdot\mid\thetabold)$, $\Delta: \mathcal{Y}^T \times \mathcal{Y}^T \rightarrow \mathbb{R}^{+}$ is a discrepancy function that measures the difference between the synthetic data set $\xbold$ and the observed data set $\ybold$. The function $K_\varepsilon(\cdot)$ is defined as $K_\varepsilon(\cdot) = \frac{K(\cdot/\varepsilon)}{K(0)}$, where $K(\cdot)$ is a centered smoothing kernel density. 
 {In this article, we mainly focus on a kernel function $K(z)$ that takes its maximum value at zero and is zero for $z > 1$. For example, the uniform, Epanechnikov, tricube, triangle,  quartic (biweight), triweight and quadratic kernels~{\citep{epanechnikov1969non, cleveland1988locally, altman1992introduction}}.}
The parameter $\varepsilon > 0$ serves as a tolerance level, and it weights the posterior $\pi_\varepsilon(\thetabold,\xbold\mid\ybold)$ higher in regions where $\xbold$ and $\ybold$ are similar.

The goal of approximate Bayesian computation is to estimate the marginal posterior distribution of $\boldsymbol{\theta}$, denoted as $\pi_{\varepsilon}(\boldsymbol{\theta} \mid \ybold)$, by integrating out the auxiliary data set $\xbold$ from the modified posterior distribution. This is given by the following equation:

\begin{equation}
\label{eq:1}
\pi_{\varepsilon}(\boldsymbol{\theta} \mid \ybold) = \frac{\pi(\boldsymbol{\theta}) \int p(\xbold \mid \boldsymbol{\theta}) K_\varepsilon(\Delta(\xbold, \ybold)) d\xbold}{\iint \pi(\boldsymbol{\theta}) p(\xbold \mid \boldsymbol{\theta}) K_\varepsilon(\Delta(\xbold, \ybold))d\xbold d\boldsymbol{\theta}}.
\end{equation}
 {The ABC posterior provides an approximation of the true posterior distribution when the tolerance level $\varepsilon$ approaches to zero.}

In practice, we set $\Delta(\xbold,\ybold)=\rho(\boldsymbol{S}(\xbold),\boldsymbol{S}(\boldsymbol{y}))$ for some distance metric $\rho(\cdot,\cdot)$ and a chosen vector of summary statistics $\boldsymbol{S}(\cdot)$: $\mathcal{Y}^T\rightarrow{\mathbb{R}^{d_s}}$  with some small $d_s$. For example, given an observed data set with $T$ independent and identically distributed data points, we can take $\boldsymbol{S}(\boldsymbol{y})=\sum_{i=1}^T \ybold_i/T$ and $\rho(\cdot,\cdot)=\left\|\cdot\right\|_2$. If the vector of summary statistic is sufficient for the parameter $\thetabold$, then comparing the summary statistics of two data sets will be equivalent to comparing the data sets themselves \citep{brooks2011handbook}. In this case, the ABC posterior converges to the conditional distribution of $\thetabold$ given $\ybold$ as $\varepsilon\rightarrow 0$; otherwise, the ABC posterior converges to the conditional distribution of $\thetabold$ given $\boldsymbol{S}(\ybold)$, that is,  {$\pi(\thetabold\mid\boldsymbol{S}(\ybold))$}. For convenience of description, we still denote $\pi_\varepsilon(\thetabold\mid\boldsymbol{S}(\ybold))$ as $\pi_\varepsilon(\thetabold\mid\ybold)$.


\subsection{ {ABC Markov chain Monte Carlo}}
 \cite{Marjoram2003MCMC} proposed an ABC Markov chain Monte Carlo (MCMC) 
method based on the Metropolis–Hastings MCMC
algorithm. 
The ABC-MCMC constructs a Markov chain that admits the target probability distribution as stationarity under
mild regularity conditions.
At $(n+1)$-th MCMC iteration, we repeat the following procedure. 
\begin{enumerate}[(i)]
\item Generate a candidate value $\thetabold^*\sim q(\cdot\mid\thetabold_{n})$.
\item Generate a synthetic data set $\boldsymbol{x}^{*}=(\xbold_{(1)}^*,\cdots,\xbold_{(T)}^*)^T$ from the likelihood, $p(\cdot\mid\boldsymbol{\thetabold^*})$.
\item Accept the proposed parameter value $(\thetabold_{n+1},\xbold_{n+1})=(\thetabold^*,\xbold^*)$ with probability
\begin{equation}
  \alpha[(\thetabold_n,\xbold_n),(\thetabold^*,\xbold^*)] = \operatorname{min}\left\{1, \frac{\pi(\thetabold^*)q(\thetabold_n\mid \thetabold^*)}{\pi(\thetabold_n)q(\thetabold^*\mid \thetabold_n)}\frac{K_\varepsilon(\Delta(\xbold^*,\ybold))}{K_\varepsilon(\Delta(\xbold_n,\ybold))}\right\};
  \label{eq:Alpha}
\end{equation}
Otherwise, set $(\thetabold_{n+1},\xbold_{n+1})=(\thetabold_n,\xbold_n)$. 
\end{enumerate}

The MCMC algorithm targets the joint posterior distribution $\pi_{\varepsilon}(\boldsymbol{\theta},\xbold\mid\boldsymbol{y})$, with the marginal distribution of interest being the posterior distribution $\pi_{\varepsilon}(\boldsymbol{\theta}\mid\boldsymbol{y})$. The sampler generates a Markov chain sequence $\{(\boldsymbol{\theta}_n,\xbold_n)\}$ for $n \geq 0$, where $\xbold_n$ represents the synthetic data set, and $\boldsymbol{\theta}_n$ represents the parameter value. 
\label{sec:ejmcmc}
For complex statistical models that are expensive to simulate, we aim to speed up inference by reducing the number of simulations.  {Based on a uniform kernel $K_\varepsilon(\Delta(\xbold,\ybold))=\mathbb{I}(\Delta(\xbold,\ybold)\leq \varepsilon)$,} \cite{Picchini2014SDE} proposed an early-rejection MCMC algorithm denoted by OejMCMC.  {Here, we generalize the uniform kernel to a more general kernel function.}
By the definition of the kernel function in Eq.\eqref{eq:1}, the acceptance probability $\alpha[(\thetabold_n,\xbold_n),(\thetabold^*,\xbold^*)]$ (Eq. \eqref{eq:Alpha}) is always not greater than
\begin{equation}
\label{eq:oejalpha}
\breve{\alpha}[(\thetabold_n,\xbold_n),\thetabold^*]=\pi(\thetabold^*)q(\thetabold_n\mid \thetabold^*)/\{\pi(\thetabold_n)q(\thetabold^*\mid\thetabold_n)K_\varepsilon(\Delta(\xbold_n,\ybold))\}.
\end{equation}
Then we can reject some candidate parameters before simulations by generating $w\sim \mathcal{U}(0,1)$, and comparing $w$ with $\breve\alpha[(\thetabold_n,\xbold_n),\thetabold^*]$. Specifically, If $w > \breve\alpha[(\thetabold_n,\xbold_n),\thetabold^*]$, we immediately reject $\boldsymbol{\theta}^*$ before simulating synthetic data with $\boldsymbol{\theta}^*$, otherwise, simulate synthetic data and determine whether to receive candidate parameters according to Metropolis-Hastings acceptance probability.  

In many cases, we may take a weak prior distribution or take a uniform distribution as prior, and use a symmetric proposal distribution, that is, for all $\boldsymbol{\theta},\boldsymbol{\theta}^*\in\Theta$,  $\pi(\boldsymbol{\theta})\approx\pi(\boldsymbol{\theta}^*)$ and $q(\boldsymbol{\theta}^*\mid\boldsymbol{\theta})=q(\boldsymbol{\theta}\mid\boldsymbol{\theta}^*)$, $P(w>\breve\alpha[(\thetabold,\xbold),\thetabold^*] )\approx 0$. Hence,  {parameters are rarely rejected early}, the method basically degenerates to the original ABC-MCMC. In Section \ref{sec:ejm}, we will  propose a new early rejection method based on discrepancy models.  

\subsection{ {ABC Sequential Monte Carlo}}

\cite{Sisson2007SMC, del2012adaptive} proposed  ABC-SMC methods within the SMC framework of \cite{del2006sequential}. 
SMC is a class of importance sampling based methods for generating weighted particles from a target distribution $\pi$ by approximately simulating from a sequence of related probability distribution $\{\pi_r\}_{r=0:R}$, where the selected initial distribution $\pi_0$ is easy to approximate (\eg the prior distribution), and the final distribution $\pi_R=\pi$ is the target distribution. The particles are moved from $r$-th target to $(r+1)$-th target, by using a Markov kernel. 

In ABC-SMC, we select a sequence of intermediate target distributions $\{\pi_{r}=\pi_{\varepsilon_{r}}(\boldsymbol{\theta}\mid \ybold)\}_{r = 0:R}$ with a descending sequence of tolerance parameters $\{\varepsilon_{r}\}_{r=0:R}$. The initial distribution has a large tolerance and is easy to approximate, the final distribution has a small tolerance and is hence accurate.
 Let $\thetabold_{r}^{(i)}$ denote the $i$-th particle after iteration $r$, let $\Delta_{r}^{(i)}$ denote the corresponding discrepancy and let $W_{r}^{(i)}$ denote the normalized weight. The set of weighted particles $\{\thetabold_{r}^{(i)},W_{r}^{(i)}\}_{i=1:N}$ represents an empirical approximation of $\pi_{\varepsilon_{r}}(\boldsymbol{\theta}\mid \ybold)$. 
 At each iteration $r+1$, we conduct propagation, weighting and resampling to approximate  $\pi_{\varepsilon_{r+1}}(\boldsymbol{\theta}\mid \ybold)$.  {We refer reader to Appendix Section B.1 for more details of these three steps. }

\section{An early rejection method based on discrepancy models}
\label{sec:ejm}

To reduce the computational cost of generating synthetic data sets in complex statistical models, we propose an early rejection method based on a discrepancy model. In Section \ref{sec:ejmcmc}, we discussed how the early rejection acceptance probability $\breve{\alpha}[(\boldsymbol{\theta}_n,\xbold_n),\boldsymbol{\theta}^*]$ proposed by \cite{Picchini2014SDE} can be  {inefficient} in certain cases. In this section, we introduce a more efficient early rejection ABC method that utilizes a discrepancy model to speed up inference. We refer to our method as {ejMCMC} to distinguish it from the early rejection MCMC proposed in \cite{Picchini2014SDE}.

\subsection{Early rejection MCMC based on a discrepancy model}
The main computational cost of ABC-MCMC comes from evaluating the acceptance ratio (Eq. \ref{eq:Alpha}), which requires simulating synthetic data sets.  {In this article,  we propose to construct a pseudo MH acceptance probability 
\begin{equation}
\label{eq:hat-alpha}
\hat{\alpha}[(\boldsymbol{\theta},\xbold),(\boldsymbol{\theta}^*,\xbold^*)]=\min\left\{1,\frac{\pi(\boldsymbol{\theta}^*) q(\boldsymbol{\theta}\mid \boldsymbol{\theta}^*)}{\pi(\boldsymbol{\theta}) q(\boldsymbol{\theta}^*\mid \boldsymbol{\theta})}
\frac{\min\{K_\varepsilon(\Delta(\xbold^*,\ybold)),K_\varepsilon(h(\thetabold^*))\}}{\min\{K_\varepsilon(\Delta(\xbold,\ybold)),K_\varepsilon(h(\thetabold))\}}\right\}
\end{equation} based on a discrepancy function $h$.}
Note that the pseudo MH acceptance probability $\hat{\alpha}[(\boldsymbol{\theta},\xbold),(\boldsymbol{\theta}^*,\xbold^*)]$ is not greater than 
\begin{equation}
\label{eq:tilde-alpha}
\tilde{\alpha}[(\boldsymbol{\theta},\xbold),\boldsymbol{\theta}^*]=\frac{\pi(\boldsymbol{\theta}^*) q(\boldsymbol{\theta}\mid \boldsymbol{\theta}^*)}{\pi(\boldsymbol{\theta}) q(\boldsymbol{\theta}^*\mid \boldsymbol{\theta})}
\frac{K_\varepsilon(h(\thetabold^*))}{\min\{K_\varepsilon(\Delta(\xbold,\ybold)),K_\varepsilon(h(\thetabold))\}},
\end{equation} 
which does not depend on $\Delta(\xbold^*,\ybold)$.  This allows us to reject some candidate parameters based on $\tilde{\alpha}[(\boldsymbol{\theta},\xbold),\boldsymbol{\theta}^*]$ before simulating the synthetic data. 
 {In the first stage of ejMCMC, we use $\tilde{\alpha}[(\boldsymbol{\theta},\xbold),\boldsymbol{\theta}^*]$ to early reject samples. 
Since $\tilde{\alpha}[(\boldsymbol{\theta},\xbold),\boldsymbol{\theta}^*]\leq \breve{\alpha}[(\thetabold_n,\xbold_n),\thetabold^*]$, ejMCMC’s early rejection rate is always not lower than OejMCMC. Our ejMCMC algorithm degenerates to OejMCMC in case that $h(\thetabold)\equiv 0$. The proposed early rejection MCMC algorithm  is shown in Algorithm \ref{alg:ejMCMC}.}

 {If $h(\thetabold)$ satisfies that $h(\thetabold)\leq \Delta(\xbold,\ybold)$ in the ABC posterior region with positive support, the pseudo MH acceptance probability (Eq.\eqref{eq:hat-alpha}) is equal to the Metropolis-Hastings acceptance probability ${\alpha}[(\boldsymbol{\theta},\xbold),(\boldsymbol{\theta}^*,\xbold^*)]$ (Eq.\eqref{eq:Alpha}). However, the function $h(\thetabold)$ cannot strictly satisfy $h(\thetabold)\leq \Delta(\xbold,\ybold)$ in practice. Consequently, we will false reject samples in region $\{\thetabold: h(\thetabold)> \Delta(\xbold,\ybold)\}$.
Hence, the ejMCMC algorithm targets a modified posterior distribution based on a discrepancy function $h(\thetabold)$, which may not be equal to the original posterior distribution $\pi_{\varepsilon}(\boldsymbol{\theta},\xbold\mid\ybold)$.}

  {As we discussed above, the function $h(\thetabold)$ is 
a trade-off between speed and accuracy. In this article, we use a Gaussian process to model the functional relationship between the discrepancy and the parameters, and $h(\thetabold)$ is $a$ (for example, $a=0.05$) quantile prediction function of the deviation. More details of the selection and training of $h(\thetabold)$ will be described in Section \ref{sec: seldm}.
 }

\begin{algorithm}[ht]
\caption{Early rejection MCMC algorithm based on a discrepancy model}
\label{alg:ejMCMC}
{\bf Input:} (a) Total number of MCMC iterations $N$, ABC threshold $\varepsilon$; (b) observed data $\ybold$; (c) prior distribution $\pi(\cdot)$, likelihood function $p(\cdot\mid \cdot)$, proposal distribution $q(\cdot\mid \cdot)$ and prediction function $h(\cdot)$.
\begin{algorithmic}[1]
\STATE Initialize $\thetabold_0$.
\FOR{$n=0,\ldots,N-1$}
\STATE Set $(\thetabold_{n+1},\xbold_{n+1})=(\thetabold_{n},\xbold_{n})$.
\STATE  Generate a candidate value $\thetabold^*\sim q(\cdot\mid\thetabold_{n})$ from the proposal distribution. 
\label{alg:ejmcmc-prop}
 \STATE Generate $w$ from $U[0,1]$.
 \IF{$w<\frac{\pi(\boldsymbol{\theta}^*) q(\boldsymbol{\theta}_{n}\mid \boldsymbol{\theta}^*)}{\pi(\boldsymbol{\theta}_{n}) q(\boldsymbol{\theta}^*\mid \boldsymbol{\theta}_{n})\min\{K_\varepsilon(\Delta(\xbold_n,\ybold)),K_{\varepsilon}(h(\thetabold_n))\}}$}
 \STATE Compute $h(\thetabold^*)$.
 \IF{$w<\tilde{\alpha}[(\boldsymbol{\theta}_n,\xbold_n),\boldsymbol{\theta}^*]$}
 \STATE generate $\xbold^*$ from $ p(\cdot\mid\boldsymbol{\theta}^*)$.
 \IF{$w<\hat{\alpha}[(\boldsymbol{\theta}_n,\xbold_n),(\boldsymbol{\theta}^*,\xbold^*)]$}
 \STATE $(\thetabold_{n+1},\xbold_{n+1})=(\thetabold^*,\xbold^*)$.
 \ENDIF
 \ENDIF
 \ENDIF
\ENDFOR
\label{alg:ejmcmc-incre}
\RETURN $\theta_{1:N}$
\end{algorithmic}
\end{algorithm}

\subsection{Properties of ejMCMC algorithm}
Let $\Theta':=\{\thetabold:h(\thetabold)<\varepsilon\}$ and  $\Theta_1:=\{\thetabold:\pi_\varepsilon(\thetabold\mid\ybold)>0\}$.  {Here $\Theta_1$ denotes the ABC posterior region with positive support}. 
The following proposition shows that under mild assumptions, the ejMCMC algorithm  satisfies the detailed balance condition. 
\begin{assumption}
\label{assum:1} 
  $\Theta'\cap\Theta_1$ is not a measure zero set.
\end{assumption}

This is a basic assumption for ejMCMC algorithm, without which the algorithm will collapse. All conclusions about the ejMCMC algorithm are based on this assumption.
\begin{proposition}
\label{prop:detail balance}
 If Assumption \ref{assum:1} is satisfied,  the
early rejection MCMC based on a discrepancy model satisfies the detailed balance condition, that is, 
{ $$\hat{\pi}_{\varepsilon}(\boldsymbol{\theta},\xbold\mid \ybold)q(\boldsymbol{\theta}^*\mid \boldsymbol{\theta})p(\xbold^*\mid\thetabold^*)\hat\alpha[(\boldsymbol{\theta},\xbold),(\boldsymbol{\theta}^*,\xbold^*)]=\hat{\pi}_{\varepsilon}(\boldsymbol{\theta}^*,\xbold^*\mid \ybold)q(\boldsymbol{\theta}\mid \boldsymbol{\theta}^*)p(\xbold\mid\thetabold)\hat\alpha[(\boldsymbol{\theta}^*,\xbold^*),(\boldsymbol{\theta},\xbold)],$$}
where \begin{equation*}
\hat{\pi}_\varepsilon(\thetabold,\xbold\mid\ybold)=\frac{\pi(\thetabold) p(\xbold\mid\thetabold)\min\{K_\varepsilon(\Delta(\xbold,\ybold)),K_\varepsilon(h(\thetabold))\}}{\iint\pi(\thetabold) p(\xbold\mid\thetabold)\min\{K_\varepsilon(\Delta(\xbold,\ybold)),K_\varepsilon(h(\thetabold))\}d\xbold d\thetabold}
\end{equation*}
 {is the invariant distribution of Markov chain.}
The marginal posterior distribution of $\thetabold$ obtained from the  ejMCMC algorithm is 
\begin{equation}
    \label{eq:posterior-ej}
\hat{\pi}_\varepsilon(\thetabold\mid\ybold)=\frac{\pi(\thetabold) \int p(\xbold\mid\thetabold)\min\{K_\varepsilon(\Delta(\xbold,\ybold)),K_\varepsilon(h(\thetabold))\}d\xbold}{\iint\pi(\thetabold) p(\xbold\mid\thetabold)\min\{K_\varepsilon(\Delta(\xbold,\ybold)),K_\varepsilon(h(\thetabold))\}d\xbold d\thetabold}.
\end{equation} 
\end{proposition}

Proposition \ref{prop:detail balance} demonstrates that using the ejMCMC move does not violate the detailed balance condition of the Markov process. Moreover, Proposition \ref{prop:accuracy} shows that the relationship between the posterior of the ejMCMC and that of the original ABC-MCMC. 

\begin{proposition}
\label{prop:accuracy} 
\begin{enumerate}[(i)]
\item  {The discrepancy function $h(\thetabold)$ satisfies that 
$$\sup_{\thetabold\in{\Theta_1/\mathcal{N}_{\Theta_1}}}\int p(\xbold\mid\thetabold)\mathbb{I}(\Delta(\xbold,\ybold)\le h(\thetabold))d\xbold\leq  a,$$ where $\mathcal{N}_{\Theta_1}$ is a Lebesgue zero-measure subset of $\Theta_1$ and $a\in[0,1)$ is a constant.} 
 The  {target} posterior distribution of ejMCMC approaches to that of  {ABC-MCMC} in terms of $L_1$ distance when $ a\to 0$, where the $L_1$-distance between $g(\cdot)$ and $f(\cdot)$  is defined as
\begin{equation}
\label{eq:L_1}
    D_{L_1}(g,f)=\int|g(\thetabold)-f(\thetabold)|d\thetabold.
\end{equation}
\item  If $K(\cdot)$ is a uniform kernel, 
the posterior distribution of ejMCMC is 
$$
    \hat{\pi}_{\varepsilon}(\boldsymbol{\theta}\mid \ybold)\propto \left\{\begin{array}{lcl}
    \pi(\thetabold)\int_{\Delta(\xbold,\ybold)<\varepsilon}p(\xbold\mid\thetabold)d\xbold&&\thetabold\in\Theta',\\
    0&&\thetabold\in\Theta/\Theta'.
\end{array}\right.$$
 {It is equal to ABC-MCMC posterior if $\Theta_1/\Theta'$ is a measure zero set.}
\end{enumerate}
\end{proposition}

Proposition \ref{prop:accuracy} shows that the resulting Markov chain still converges to the true approximate Bayesian posterior under some conditions of $h(\cdot)$. This implies that the {ejMCMC} algorithm provides a promising framework to Bayesian posterior approximation, without compromising on accuracy or convergence guarantees  {under some mild conditions}. In some other cases, there may exist some mismatch between the posterior of ejMCMC and ABC-MCMC. In this article, we utilize the $L_1$-distance shown in Eq. \eqref{eq:L_1} to quantify the dissimilarity between two density functions. 

 {Next we will prove the posterior concentration
property of ejMCMC algorithm.} \cite{frazier2018asymptotic} proved the posterior concentration property of the ABC posterior with the uniform kernel, that is, for any $\delta>0$, and for some $\varepsilon>0$,
$$\Pi_{\varepsilon}\{d(\thetabold,\thetabold_0)>\delta\mid \ybold\}=\int_{d(\thetabold,\thetabold_0)>\delta}\pi_\varepsilon(\thetabold\mid \ybold)d\thetabold=o_{P}(1),$$ where $d(\cdot,\cdot)$ is a metric on $\Theta$. Theorem \ref{th:concentration-ej} shows that the posterior concentration of the algorithm still holds, with some technical assumptions of function $h(\cdot)$. This property is important, because for any $A\subset\Theta$,  {ABC posterior probability} $\Pi_{\varepsilon}\{A\mid \ybold\}$ will be different from the exact posterior probability. Without the guarantees of exact posterior inference, knowing that $\Pi_{\varepsilon}\{\cdot \mid \ybold\}$ will concentrate on  {the true parameter} $\thetabold_0$ can be used as an effective way to express our uncertainty about $\thetabold$. The posterior concentration is related to the rate at which $\varepsilon$ goes to 0 and the rate at which the observed summaries $S(\ybold)$ and the simulated summaries $S(\xbold)$ converge to well defined limit counterparts $b(\thetabold_0)$ and $b(\thetabold)$. Here we consider $S(\xbold)\rightarrow b(\thetabold)$ as $T\rightarrow\infty$, where $\xbold=(\xbold_{(1)},\ldots,\xbold_{(T)})^\top$ and $\xbold_{(t)}\sim p(\cdot\mid\thetabold)$ for $t=1,\ldots,T$, and consider $\varepsilon$ as a $T$-dependent sequence satisfying $\varepsilon_T\rightarrow 0$ as $T\rightarrow \infty$.  {Here, $\Pi(\cdot)$ is a probability measure with prior density $\pi(\cdot)$.}

\begin{assumption}
    \label{assum:2}
    For all $\varepsilon$, there exists a constant $0\le\gamma_1< 1$ satisfying that $$\Pi\left[\left\{\thetabold:h(\thetabold)\ge\varepsilon\text{ and } \rho(b(\thetabold),b(\thetabold_0))\leq{\varepsilon/3}\right\}\right]\le\gamma_1\Pi\left[\rho(b(\thetabold),b(\thetabold_0))\leq{\varepsilon/3}\right].$$
\end{assumption}
\begin{assumption}
\label{assum:3}
    For all $\varepsilon$, there exists a constant $0\le\gamma_2< 1$ satisfying that
    $$\iint_{\Theta'} p(\xbold\mid\thetabold)\mathbb{I}(\rho(S(\xbold),S(\ybold))< h(\thetabold))d\xbold d\Pi(\thetabold) \le \gamma_2\iint_{\Theta'} p(\xbold\mid\thetabold)\mathbb{I}(\rho(S(\xbold),S(\ybold))< \varepsilon)d\xbold d\Pi(\thetabold). $$
\end{assumption}
\textit{Remark.}  {Assumption \ref{assum:2} controls the proportion of $\thetabold$ satisfying $h(\thetabold)\geq \varepsilon$, that will be rejected before simulating, in a neighbourhood of $\thetabold_0$.  Assumption \ref{assum:3} is required for ABC with a more general kernel than a uniform kernel. This assumption controls the proportion of acceptable $(\thetabold,\xbold)$ that may be rejected early. 
Intuitively, Assumptions \ref{assum:2} and \ref{assum:3} about $h(\thetabold)$ are to ensure that a small proportion of $\thetabold$ near $\thetabold_0$ will be rejected before simulating. As detailed in Section \ref{subsec:dis model}, our proposed discrepancy model $h(\thetabold)$ can always ensure that only a small proportion of $\thetabold$ near the true value are early rejected by adjusting the quantile prediction function. Without these assumptions about the early rejection stage,  posterior concentration concentration about $\thetabold_0$ cannot occur.
}


\begin{theorem}
\label{th:concentration-ej}
     If  Assumptions 1-3 of \cite{frazier2018asymptotic} and Assumptions \ref{assum:1}-\ref{assum:3} stated in this article are satisfied, the posterior concentration of the ejMCMC targeting Equation \eqref{eq:posterior-ej} still holds. For uniform kernel cases, Assumption \ref{assum:3} is not required for this property.
\end{theorem}

We have previously discussed the quality of the ABC estimates. Now, we turn to investigate the efficiency of the algorithm.

\begin{proposition}
\label{prop:efficiency}  

When MCMC algorithms reach the stationary distributions, the early rejection rates of ejMCMC algorithm and OejMCMC are 
$$R_{ej}=1-\iint\min\{1,\tilde{\alpha}[(\boldsymbol{\theta},\xbold),\thetabold^*]\}q(\thetabold^*\mid\thetabold)\hat{\pi}_\varepsilon(\thetabold,\xbold\mid\ybold)d\xbold d\thetabold^*d\thetabold$$
and
$$R_{Oej}=1-\iint\min\{1,\Breve{\alpha}[(\boldsymbol{\theta},\xbold),\thetabold^*]\}q(\thetabold^*\mid\thetabold){\pi}_\varepsilon(\thetabold,\xbold\mid\ybold)d\xbold d\thetabold^*d\thetabold,$$
respectively. For uniform kernel cases, if $\Theta_1-\Theta'$ is a set of measure zero on parameter space $\Theta$, 
\begin{equation}
    R_{ej}-R_{Oej}\ge \int_{\Theta'\cap\Theta_1}\int_{\Theta/\Theta'}\min\{1,\frac{\pi(\thetabold^*)q(\thetabold\mid\thetabold^*)}{\pi(\thetabold)q(\thetabold^*\mid\thetabold)}\}q(\thetabold^*\mid\thetabold)\pi_\varepsilon(\thetabold\mid\ybold)d\thetabold^*d\thetabold,
    \label{eq:efficiency}
\end{equation} 
where $\Theta'=\{\thetabold:h(\thetabold)<\varepsilon\}$ and $\Theta_1=\{\thetabold:\pi_\varepsilon(\thetabold\mid\ybold)>0\}$.
\end{proposition}

Proposition \ref{prop:efficiency} shows that when the kernel function $K(\cdot)$ is uniform and the posterior of ejMCMC and OejMCMC are consistent, the efficiency of ejMCMC is expected to be higher since the term $R_{ej}-R_{Oej}$ in Equation \eqref{eq:efficiency} is always positive. This is because ejMCMC targets the posterior region more effectively by rejecting more candidates in the early rejection step, potentially reducing the computational cost of generating synthetic data sets.  {Furthermore, when the function $h(\thetabold)$ satisfies $h(\thetabold) > \varepsilon$ in all regions with negligible posterior probabilities, the efficiency of ejMCMC is maximized. This means that the discrepancy model $h(\thetabold)$ is able to identify high probability ABC posterior regions, leading to higher early rejection rates and improved efficiency of the algorithm.} The ejMCMC algorithm can play a role in correcting prior information before simulations in some sense, when the prior information is poor. In this article, we use 
\begin{equation}
    \text{Eff}=\frac{\text{\# of early rejected parameters}}{\text{\# of rejected parameters}}
    \label{eq:Eff}
\end{equation} to quantify the early rejection efficiency of algorithms. This metric can be seen as a regularization of the early rejection rate, ensuring that the efficiency value is bounded between 0 and 1, with 1 being the best possible efficiency where all rejected parameters are early rejected. Proposition \ref{prop:EFF} shows the effect of the function $h(\cdot)$ on the early rejection efficiency of the ejMCMC algorithm.
\begin{proposition}
\label{prop:EFF}
For all $\thetabold\in\Theta$,  larger probability of $\Delta(\xbold,\ybold)\le h(\thetabold)$ can lead to  higher early rejection efficiency of ejMCMC. 
\end{proposition}
 Combining Propositions \ref{prop:accuracy} and \ref{prop:EFF}, the probability of $\Delta(\xbold,\ybold)\le h(\thetabold)$ keeps a balance between efficiency and accuracy.
\subsection{An early rejection adaptive SMC}
\label{subsec:ejSMC}
 {
ABC-MCMC algorithm has several drawbacks. Firstly, the convergence speed of MCMC algorithm is often slow, especially in high-dimensional spaces. This may require a large number of iterations to obtain accurate approximations. Secondly, the MCMC algorithm is sensitive to the choice of initial values. Improper initial values can cause the Markov chain to get stuck in local optima, preventing effective exploration of the entire state space. 
In comparison with ABC-MCMC, ABC-SMC employs a sequential updating of particle weights to approximate the target distribution, which effectively mitigates the risk of getting stuck in local optima. Furthermore, the parallelization in SMC algorithms enhance their efficiency in complex applications.}
In this section, we introduce an early rejection ASMC ({ejASMC}) based on ABC adaptive SMC (ABC-ASMC) \citep{del2012adaptive} and the {ejMCMC} method.
The sequence of thresholds and MCMC kernels can be pre-specified before running the algorithm, but  the algorithm may collapse with improperly selected thresholds and kernels. 
 In this article, we adaptively tune the sequence of thresholds and {ejMCMC} proposal distributions.

\vspace{0.5em}
\textbf{Adaptive selection of thresholds:} We adapt the scheme proposed in \cite{del2012adaptive} for selecting the sequence of thresholds, based on the key remark that the evaluation of weights  {$W_{r+1}^{(i)}={W}_{r}^{(i)}K_{\varepsilon_{r+1}}
(\Delta(\ybold,\xbold_{\thetabold_{r}^{(i)}})/K_{\varepsilon_r}(\Delta(\ybold,\xbold_{\thetabold_{r}^{(i)}}))$} does not depend on particles of iteration $r+1$, $\{\thetabold_{r+1}^{(i)}, \xbold_{r+1}^{(i)}\}_{i=1:N}$. We select the threshold by controlling the proportion of unique alive particles, which is defined as 
\begin{equation*}
    \operatorname{PA}(\{W_{r+1}^{(i)}\},\varepsilon_{r+1})=\frac{\sum_{i\in U}\mathbb{I}(W_{r+1}^{(i)}>0)}{N},
\end{equation*}
where $W_{r+1}^{(i)}$ is a function of the threshold $\varepsilon_{r+1}$, and $U$ is the largest subset of the set $\{1:N\}$ such that for any two distinct elements $i$ and $j$ in $U$, $\thetabold_{r+1}^{(i)}\neq \thetabold_{r+1}^{(j)}$. The proportion of unique 
alive particles is also intuitively a sensible measure of ‘quality’ of our SMC approximation.
The threshold $\varepsilon_{r+1}$ is selected to make sure that the proportion of alive particles is equal
to a given percentage of the particles number \begin{equation}
\label{eq:epsilon-solve}
    \operatorname{PA}(\{W_{r+1}^{(i)}\},\varepsilon_{r+1})=\gamma N
\end{equation}
for $\gamma\in(0,1)$. In practice, bisection is used to compute the root of Equation \eqref{eq:epsilon-solve}. 
The parameter $\gamma$ is a ``quality''
index for the resulting SMC approximation of the target. If $\gamma\rightarrow 1$ then we will move
slowly towards the target but the SMC approximation is more accurate. However, if $\gamma\rightarrow 0$
then we can move very quickly towards the target but the resulting SMC approximation will be unreliable.

\vspace{0.5em}
\textbf{Adaptive MCMC proposal distribution:}
For each intermediate target, the SMC algorithm applies an MCMC move with invariant density $\pi_{\varepsilon_{r+1}}(\thetabold\mid\ybold)$. MCMC moves are implemented by accepting candidate parameters with a certain probability given by the MH ratio of ejMCMC algorithm. For $\thetabold_{r}^{(i)}$ with $w_{r+1}^{(i)}>0$, we generate $\thetabold^*,\xbold^*$ according to a proposal $q_{r+1}(\thetabold^*\mid\thetabold_{r}^{(i)})p(\xbold^*\mid\thetabold^*)$, then accept the candidate with probability $\hat{\alpha}[(\thetabold_r^{(i)},\xbold_r^{(i)}),(\thetabold^*,\xbold^*)]$. 
We can adaptively determine the parameters of the proposal $q_{r+1}(\cdot\mid\cdot)$ based on the previous approximation of the target $\pi_{\varepsilon_{r}}$. In this article, we use Gaussian distribution as the proposal distribution, and the Gaussian covariance matrix is determined adaptively by computing the covariance matrix of the weighted particles $\{\thetabold_{r}^{(i)},W_{r}^{(i)}\}_{i=1:N}$.

The output of {ejASMC} is the weighted particles $\{\thetabold_R^{(i)},W_R^{(i)}\}_{i=1:N}$, the empirical distribution of which is the ABC posterior estimate. We set $\varepsilon_0=\infty$, then the initial target distribution is the prior distribution. Thus the initial weighted particles $\thetabold_0^{(i)}\sim\pi(\thetabold)$ and $W_0^{(i)}=1/N$ for $i=1,\cdots,N$.  {The detailed pseudo-code of {ejASMC} is shown in Appendix Section B.2.}


\subsection{Selection of discrepancy model}
\label{sec: seldm}
\subsubsection{A Gaussian process discrepancy model $h(\cdot)$}
\label{subsec:dis model}
In approximate Bayesian computation, whether a candidate parameter $\boldsymbol{\theta}^*$ will be retained as a sample of the posterior distribution is mainly determined by the discrepancy $\Delta(\xbold,\boldsymbol{y})$ between the pseudo-data set $\boldsymbol{x}\sim p(\cdot\mid\boldsymbol{\theta}^*)$ and the observed data $\ybold$. To speed up the inference, \cite{gutmann2016bayesian}  {considered a  uniform kernel in Eq.\eqref{eq:1} and}  
proposed to model the discrepancy  $\Delta_{\boldsymbol{\theta}} =\Delta(\xbold_{\boldsymbol{\theta}},\ybold)$ between the observed data
$\ybold$ and the pseudo-data $\boldsymbol{x}_{\boldsymbol{\theta}}$ as a function of $\boldsymbol{\theta}$. The estimated posterior  {with a uniform kernel} for each $\boldsymbol{\theta}$ admits 
\begin{equation}
\pi_\varepsilon(\thetabold\mid\ybold)\propto
\pi(\boldsymbol{\theta})\mathbb{P}(\Delta_{\boldsymbol{\theta}}\leq\varepsilon),
\end{equation} where the probability is computed using the fitted surrogate model. Furthermore, for a continuous and strictly increasing function $g(\cdot)$, we have $\mathbb{P}(g(\Delta_{\boldsymbol{\theta}})\leq g(\varepsilon)) =\mathbb{P}(\Delta_{\boldsymbol{\theta}}\leq\varepsilon)$. In some cases, modeling $g(\Delta_{\boldsymbol{\theta}})$ instead of $\Delta_{\boldsymbol{\theta}}$ as a function of $\boldsymbol{\theta}$ is easier and more suitable. For example, as $\Delta(\ybold,\xbold_{\thetabold})$ is a positive value for all $\xbold_{\thetabold}$, it may be a better choice to model log-discrepancy $\log(\Delta(\ybold,\xbold_{\thetabold}))$.

 {\cite{gutmann2016bayesian,jarvenpaa2019Efficient,jarvenpaa2020batch} utilize GP formulations to model the discrepancy.}
In GP regression, we assume that $\Delta_{\boldsymbol{\theta}}\sim \mathcal{N}(f(\boldsymbol{\theta}),\sigma^2)$ and $f(\boldsymbol{\theta})\sim \mathcal{GP}(m(\boldsymbol{\theta}),k(\boldsymbol{\theta},\boldsymbol{\theta}'))$ with a mean function $m(\cdot)$: $\Theta\rightarrow{\mathbb{R}}$ and a covariance function $k(\cdot,\cdot)$: $\Theta\times\Theta\rightarrow{\mathbb{R}}$. For a  {training data set} $\mathcal{D}_s=\{(\Delta_1,\boldsymbol{\theta}_1),\cdots,(\Delta_s,\boldsymbol{\theta}_s)\}$, the posterior predictive density for the latent function $f$ at $\boldsymbol{\theta}$ follows a Gaussian density with the mean and variance
\begin{equation}
   \mu_{f\mid \mathcal{D}_s}(\boldsymbol{\theta})=\mathbf{k}_s(\boldsymbol{\theta})^T \mathbf{K}_s^{-1}(\boldsymbol{\theta}) \Delta_{1: s}, \quad v_{f\mid \mathcal{D}_s}(\boldsymbol{\theta})=k(\boldsymbol{\theta}, \boldsymbol{\theta})-\mathbf{k}_s(\boldsymbol{\theta})^T \mathbf{K}_s^{-1}(\boldsymbol{\theta}) \mathbf{k}_s(\boldsymbol{\theta}) 
\end{equation}
respectively. Here $\Delta_{1:s}=\left(\Delta_1,\cdots,\Delta_s\right)^T$, $\mathbf{k}_s(\boldsymbol{\theta}) = \left(k(\boldsymbol{\theta},\boldsymbol{\theta}_1),\cdots,k(\boldsymbol{\theta},\boldsymbol{\theta}_s)\right)^T$, and $\mathbf{K}_s(\boldsymbol{\theta})$ is a $s\times s$ matrix with $[\mathbf{K}_s(\boldsymbol{\theta})]_{ij}=k(\boldsymbol{\theta}_i,\boldsymbol{\theta}_j)+\sigma^2\mathbb{I}_{i=j}$ for $i=1,\ldots s$ and $j=1,\ldots s$. The ABC posterior at $\boldsymbol{\theta}$ can be computed using the fitted GP as
\begin{equation}
\pi_\varepsilon(\thetabold\mid\ybold)\propto\pi(\boldsymbol{\theta})\mathbb{P}\left(\Delta_{\boldsymbol{\theta}} \leq \varepsilon\right)=\pi(\boldsymbol{\theta})\Phi\left(\left(\varepsilon-\mu_{f\mid\mathcal{D}_s}(\boldsymbol{\theta})\right) / \sqrt{v_{f\mid\mathcal{D}_s}(\boldsymbol{\theta})+\sigma^2}\right),
\end{equation}
where $\varepsilon$ is the threshold and $\Phi$ is the cumulative density function of the standard Gaussian distribution.  



Following \cite{gutmann2016bayesian} and \cite{Jarvenpaa2018GPABC,jarvenpaa2019Efficient,jarvenpaa2020batch}, we use a Gaussian process model to establish the functional relationship between the discrepancy and the parameters, and propose an expression for $h(\thetabold)$.
We use an initial discrepancy-parameters pairs $\mathcal{D}_{s}$ to train a GP model. Based on the GP discrepancy model, we take 
\begin{equation}
\label{eq:h}
    h_a(\boldsymbol{\theta})=\mu_{f\mid\mathcal{D}_{s}}(\boldsymbol{\theta})-Z_{a}\sqrt{v_{f\mid\mathcal{D}_{s}}(\boldsymbol{\theta})+\sigma^2},
\end{equation}
where $\mathbb{P}(x<Z_{a})=a$ and $x\sim N(0,1)$. Let $h(\boldsymbol{\theta})=h_{a}(\boldsymbol{\theta})$, Propositions \ref{prop:accuracy} and \ref{prop:EFF} indicate that the choice of $a$ is a trade-off between efficiency and accuracy of ejMCMC algorithm. A smaller value of $a$ will lead to more accurate estimation of ejMCMC algorithm but lower early rejection efficiency.

\subsubsection{Selection of training data}
\label{sec: train}

Although the posterior estimated by an early rejection method based on the discrepancy model does not rely entirely on the fitted surrogate model like GP-ABC \citep{Jarvenpaa2018GPABC}, the accuracy of the discrepancy model is also critical to the results. Proposition \ref{prop:accuracy} shows that we only need to ensure the accuracy of discrepancy model in approximation Bayesian posterior region, that is, the region with small discrepancy.
In \cite{gutmann2016bayesian,jarvenpaa2019Efficient}, they use a GP discrepancy model to intelligently select the next parameter value $\boldsymbol{\theta}_{s+1}$ via minimization the acquisition function \begin{equation}
    \mathcal{A}_s(\boldsymbol{\theta})=\mu_{f\mid\mathcal{D}_s}(\boldsymbol{\theta})-\eta_s\sqrt{v_{f\mid\mathcal{D}_s}(\boldsymbol{\theta})},
\end{equation} where $\eta_s$ is a positive number related to $s$, then run the computationally costly
simulation model to obtain updated training data $\mathcal{D}_{s+1}$. The coefficient $\eta_s$ is a trade-off between exploration and exploitation. This method of sequentially generating training points can ensure that there are enough training data points in areas with small discrepancy to ensure the accuracy of the model in this area. But it is computationally expensive to select a new training point every time by solving an optimization problem. 

In this article, we consider ensuring the accuracy of the model by placing more training points in areas with smaller discrepancy via a pilot run of ABC-SMC algorithm with {OejMCMC} as proposals to sequentially generate batch data points.  {The detailed pseudo-code is shown in Appendix Section B.3.}
Our algorithm does not require to solve the optimization problem, and hence has lower computational cost. The budgets for the pilot ABC-SMC run could be relatively small (\eg with a short sequence of tolerance parameters). Moreover, when the accuracy of the discrepancy model trained by ABC-SMC samples is not enough, the algorithm allows to continue to generate training data sequentially, 
by changing the algorithm termination conditions. 

 {The computational complexity of training a GP model is $O(n^3)$. Increasing sample size of training data will significantly increase the computational burden. Based on \cite{gutmann2016bayesian,Jarvenpaa2018GPABC,jarvenpaa2019Efficient} and our exploratory study, we suggest collecting a few thousands data points for training GP discrepancy model. However, this may not be enough for large dimensional problems. With increasing number of iterations, we can use the 
set of accepted parameters and corresponding discrepancy terms as new training data to update the GP
model within ejMCMC \citep{drovandi2018accelerating}.
In this case, we recommend using the online-GP algorithm \citep{stanton2021onlineGP} to train the Gaussian process model, which can use new training data to update the model without using the full data sets.} 

\section{Simulation study}
\label{sec:sim}

\subsection{A toy example}
\label{subsec:toy}
We first illustrate the advantage of our proposed method via a toy example. Suppose that the model of interest is a mixture of two normal distributions
$$p(y\mid \theta)=0.5\phi(y;\theta+2,0.6)+0.5\phi(y;\theta -1,0.6),$$
where $\phi(y;\mu,\sigma^2)$ denotes the probability density function of $\mathcal{N}(\mu,\sigma^2)$ evaluated at $y$. Suppose that the prior distribution is $\pi(\theta)=\mathcal{U}(-6,6)$, and the observed data associated with this model is $y_0 = 1$. The discrepancy we use is $\Delta(y,y_0)=|y-1|$. We choose  {a uniform kernel} with the threshold $\varepsilon=0.6$ for ABC algorithms.  {Note that we only have one observed data point in this example.}

\begin{figure}[h]
\centering  
\subfigure[]{   
\begin{minipage}{0.45\textwidth}
\centering   
\includegraphics[width=0.95\linewidth]{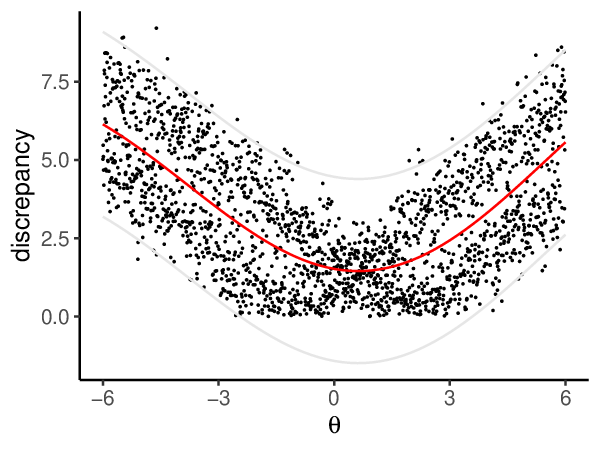}
\label{fig:GM1_GP}
\end{minipage}
}
\subfigure[]{
\begin{minipage}{0.45\textwidth}
\centering   
\includegraphics[width=0.95\linewidth]{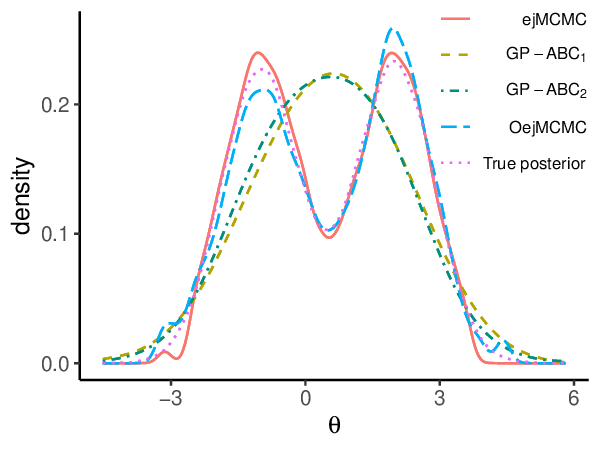}
\label{fig:GM1}
\end{minipage}
}
\caption{(a) Gaussian process modeling for identifying the relationship between the discrepancy $\Delta$ and parameter $\theta$. The x-axis denotes the parameter and the y-axis  denotes the value of the discrepancy. 
Black dots denote the discrepancies, the red line is the mean function of GP and the area between the two grey lines is the 95\% predictive interval. (b)   {Posterior density estimated by GP-ABC$_1$, GP-ABC$_2$, {ejMCMC}, OejMCMC, and rejection sampling (ground truth). }  }
\end{figure}

Firstly, we run rejection sampling ABC $10^6$ iterations and treat the resulting posterior distribution as ground truth.
  {We then compare our ejMCMC with OejMCMC and GP-ABC \citep{Jarvenpaa2018GPABC}.}
  {We collect $2,000$ training data points from prior, which are then used to train the GP discrepancy model. The resulting GP model is used to approximate the posterior of GP-ABC, and we name this as GP-ABC$_1$.
 We then run GP-ABC, {OejMCMC} and {ejMCMC} with same computational budgets ($N = 100,000$ iterations). This GP-ABC with $N = 100,000$ is named as GP-ABC$_2$.}
A random walk proposal $q(\theta\mid\theta_{n})\sim\mathcal{N}(\theta_{n},0.3^2)$ is implemented for MCMC algorithms. 
In the implementation of {ejMCMC}, the discrepancy model $h_a(\theta)$ is a GP model trained  with $a=0.05$. 
Figure \ref{fig:GM1_GP} shows the Gaussian process modeling for identifying the relationship between the discrepancy $\Delta$ and parameter $\theta$. The x-axis denotes the parameter and the y-axis  denotes the value of the discrepancy. 
Black dots denote the discrepancies, the red line is the mean function of GP and the area within two grey lines is the 95\% predictive interval. In this example, the discrepancy for parameters is multi-modal, GP discrepancy model doesn’t fit the `shape' of the discrepancy correctly.   {Figure \ref{fig:GM1}
displays the estimated posterior distributions provided by  
GP-ABC$_1$, GP-ABC$_2$, {ejMCMC}, OejMCMC and the ground truth (rejection sampling). It indicates that both GP-ABC algorithms fail to capture a bimodal shape of the posterior, due to that they fail to capture the bimodal shape of discrepancy.} This is also demonstrated in \cite{Jarvenpaa2018GPABC}. 
Our {ejMCMC} is able to capture both peaks of the posterior. The $L_1$ distance ($D_{L_1}$) between the estimated posterior by {ejMCMC} and the corresponding true posterior is 0.0559, lower than the $D_{L_1}$ between {OejMCMC} and ground truth (0.0748).  {We also implement ejMCMC with online-GP \citep{stanton2021onlineGP} discrepancy model for this toy example. It is available at \url{https://github.com/caofff/ejMCMC}.}


\subsection{An ordinary differential equation model}
In this subsection, we investigate the performance of ABC algorithms via an ordinary differential equation (ODE) example. 
We use the R package \emph{deSolve} \citep{soetaert2010solving} to simulate differential equations.
We generate ODE trajectories according to the following system,
\begin{eqnarray}
\label{eq: ode1}
\frac{dx_{1}(t)}{dt} &=& \frac{72}{36+x_{2}(t)} - \theta_{1},\nonumber\\
\frac{dx_{2}(t)}{dt} &=& \theta_{2}x_{1}(t) - 1,
\end{eqnarray}
where $\theta_{1} = 2$ and $\theta_{2} = 1$, and the initial conditions $x_{1}(0) = 7$ and $x_{2}(0) = -10$. The observations $\ybold_{i}$ are simulated from a normal distribution with the mean  $x_{i}(t|\thetabold)$ and the variance $\sigma_{i}^{2}$, where $\sigma_{1} = 1$ and $\sigma_{2} = 3$. We generate $121$ observations for each ODE function, equally spaced within $[0, 60]$. We use root mean squared error (RMSE) to model the discrepancy 
between the observed data and the pseudo-data. We take the following prior distributions $\pi(\thetabold)$: $\theta_1\sim\mathcal{U}(1.8,2.2)$, $\theta_2\sim\mathcal{U}(0.8,1.2)$.

We conduct a performance comparison of two early rejection MCMC algorithms under different threshold levels. A smaller threshold value implies that the ABC posterior is closer to the actual posterior, and consequently, more simulation costs are required. In the ABC-MCMC algorithm, the convergence speed slows down with smaller thresholds. The 1\% and 5\% quantiles of the discrepancy correspond to the parameters that conform to the prior distribution are 4.03 and 4.68, respectively. Here we study the performance of ABC algorithms with $\varepsilon=4.8$, $4.5$, $4.1$ and $4.05$. We observe the Markov chains mix poor when the threshold value is smaller. 
For each level of tolerance, we first run MCMC algorithms $1,000,000$ iterations with 3 different initializations. The Gelman-Rubin diagnostic \citep{gelman1995bayesian} is used to assess the convergence of the Markov chain. The resulting posterior can be regarded as ground truth. For each case, we also run both ejMCMC and OejMCMC $100,000$ iterations.  {We collect $3,000$ training data from prior for ejMCMC algorithm. Appendix Figure C.1 illustrates the efficacy of the Gaussian Process (GP) model's fit and the positioning of the training data points. The discrepancy can be fitted well by the Gaussian process.}  The choice of $a$ in the GP discrepancy model is a trade off between efficiency and accuracy. For each threshold $\varepsilon$, we run the ejMCMC algorithm with $a=0.5$, $0.2$, $0.1$, $0.05$ and $0.01$.  {Table \ref{tab:ODEa} shows the results for $\varepsilon=4.5$ by varying $a$ ({\it i.e.} $h(\thetabold)$). This demonstrates that the selection of $h(\thetabold)$ keeps a balance between accuracy and computing speed. 
Results for the remaining three thresholds are shown in Appendix Section C.1.}
To balance the efficiency and accuracy, we choose $a=0.05$. 

\begin{table}[h]
\centering
\caption{ODE: $\varepsilon=4.5$, accuracy and efficiency of ejMCMC varying $a$.}
\begin{tabular}{cccccc}
  \hline
$a $& $N_{\text{sim}}$ & $N_{\text{acc}}$ & $D_{L_1}(\boldsymbol{\theta}_1)$ & $D_{L_1}(\boldsymbol{\theta}_2)$ &Eff \\ 
  \hline
 0.50 & 40039 & 36170 & 0.2071 & 0.1922 & 0.9394 \\ 
  0.20 & 47550 & 37574 & 0.0541 & 0.0459 & 0.8402 \\ 
 0.10 & 51398 & 37572 & 0.0374 & 0.0426 & 0.7785 \\ 
 0.05 & 55116 & 38017 & 0.0226 & 0.0216 & 0.7241 \\ 
 0.01 & 60464 & 37593 & 0.0273 & 0.0194 & 0.6335 \\ 
   \hline
\end{tabular}
\label{tab:ODEa}
\end{table}

Table \ref{tab:ode} shows the number of synthetic data generated ($N_{\text{sim}}$), number of collected posterior samples ($N_{\text{acc}}$), and the $L_1$-distance between true value and posterior estimates  varying threshold parameter $\varepsilon$.  {Under the settings of uniform prior and Gaussian proposal distribution, the efficiency of OejMCM is 0. While} our proposed algorithm can achieve a considerable computational acceleration by saving $20\%-80\%$ of data simulation. The accuracy of the posterior distribution estimates has no drop compared to {OejMCMC}. The efficiency of algorithms with threshold values $4.8$ and $4.5$ is higher than cases with  that of $4.1$ and $4.05$.  When the threshold is set to 4.1 or 4.05, the proportion of candidate parameters lying outside of the posterior region is low, resulting in a decrease of efficiency. 

\begin{table}[h]
\centering
\caption{Efficiency of OejMCMC and ejMCMC varying threshold parameter $\varepsilon$.}

\begin{tabular}{ccccccc}

  \hline
 $\varepsilon$ & method & $N_{\text{sim}}$ & $N_{\text{acc}}$ & $D_{L_1}(\boldsymbol{\theta}_1)$ & $D_{L_1}(\boldsymbol{\theta}_2)$ &Eff\\ 
  \hline
  \multirow{2}*{4.8} & {OejMCMC }  & 100000 & 39965 & 0.0207 & \textbf{0.0171 }& 0 \\ 
   & {ejMCMC} & 52719 & 39549 & \textbf{0.0194} & 0.0216 & 0.7821 \\
   \hdashline
  \multirow{2}*{4.5} & {OejMCMC }  & 100000 & 37691 & 0.0244 & 0.0259 & 0 \\ 
   & {ejMCMC} & 55116 & 38017 & \textbf{0.0226} &\textbf{ 0.0216} & 0.7241 \\ 
   \hdashline
  \multirow{2}*{4.1} & {OejMCMC }  & 100000 & 37672 & \textbf{0.0418} & 0.0452 & 0 \\ 
   & {ejMCMC}& 76538 & 38218 & 0.0427 & \textbf{0.0360 }& 0.3798 \\ 
   \hdashline
  \multirow{2}*{4.05} & {OejMCMC } & 100000 & 39861 & 0.0445 & 0.0533 & 0 \\ 
  & {ejMCMC} & 85325 & 39897 & \textbf{0.0428} & \textbf{0.0420} & 0.2442 \\ 
   \hline
\end{tabular}
\label{tab:ode}
\end{table}



\begin{figure}
    \centering
    \includegraphics[width=0.8\linewidth]{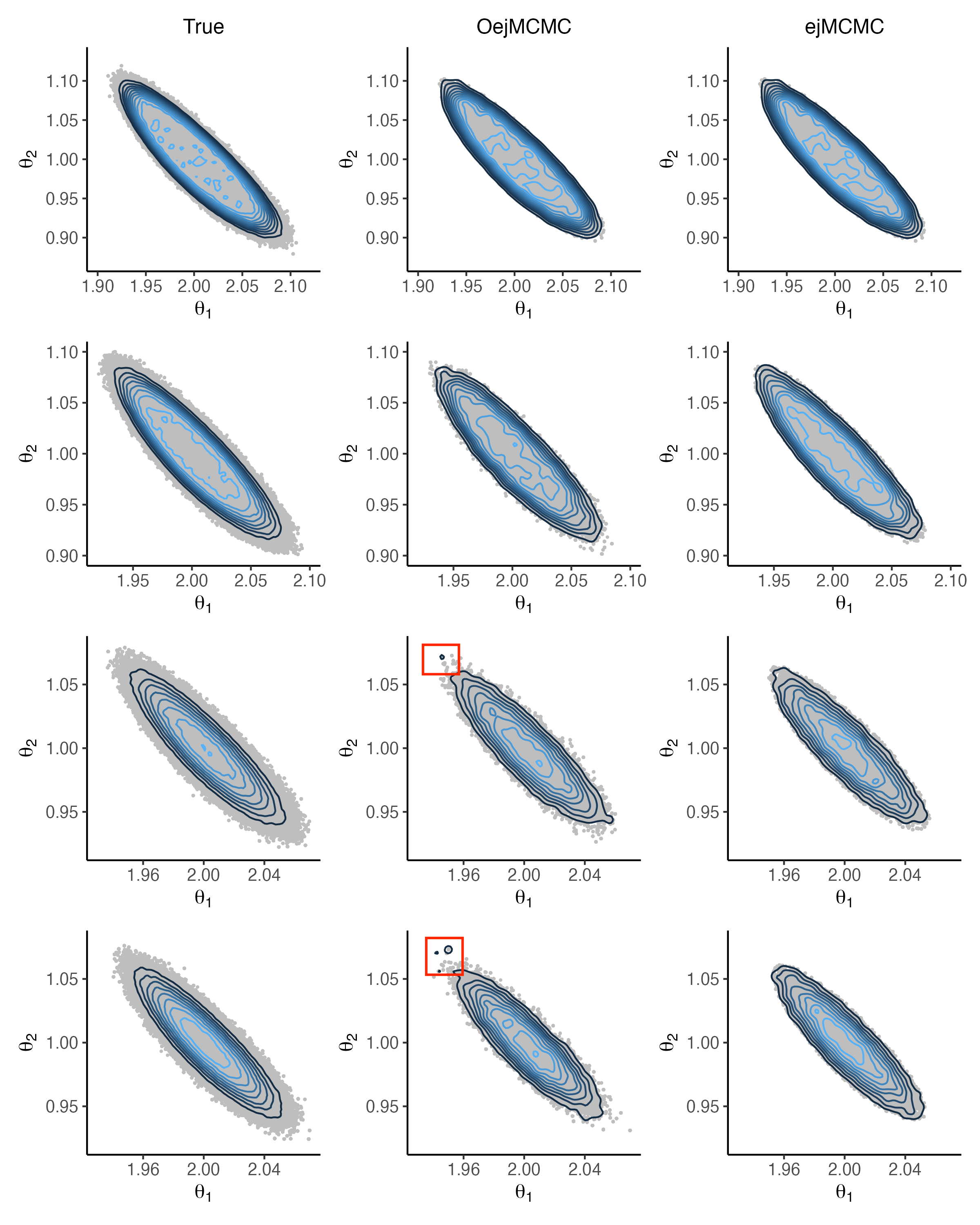}
    \caption{Posterior distributions of the ODE model provided by the three ABC algorithms varying the threshold parameter $\varepsilon$. The grey dots represent the accepted MCMC samples.  First column: the true posterior estimated by running reject sampling with
$100,000$ accepted particles. Second column: posterior density estimated by {OejMCMC} with $100,000$ iterations.
Third column: posterior density estimated by {ejMCMC} with $100,000$ iterations. First row: $\varepsilon=4.8$, second row: $\varepsilon=4.5$, third row: $\varepsilon=4.1$, last row: $\varepsilon=4.05$.}
    \label{fig:ODE:2d}
\end{figure}
Figure \ref{fig:ODE:2d} shows the posterior distributions of the ODE model provided by the three ABC algorithms varying the threshold parameter $\varepsilon$. 
The distributions provided by three methods are close, though ejMCMC slightly misses the tail of posteriors. This is because all candidate parameters satisfying $h(\thetabold)>\varepsilon$ in the posterior region are early rejected.
However, the false rejection avoids the MCMC sampler ``sticking" in some local region for long periods
of time (marked by the red box in the Figure \ref{fig:ODE:2d}). We show an example of ABC-MCMC traceplots in  {Appendix Figure C.2}.

\begin{figure}[H]
\centering  
\includegraphics[width=0.8\linewidth]{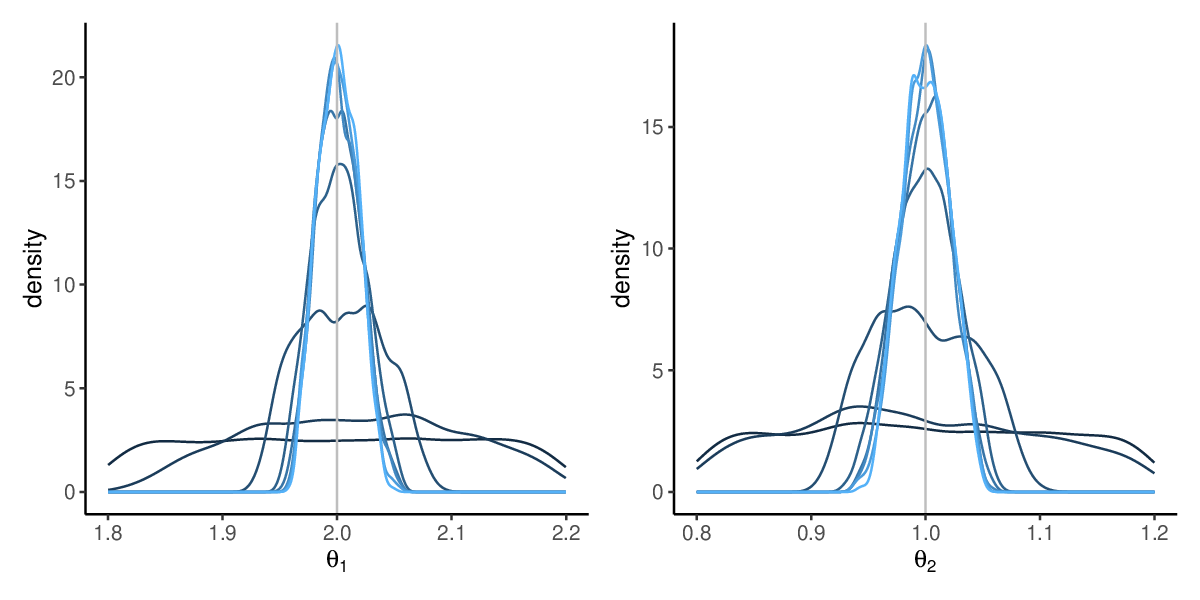}
\caption{Marginal posterior distributions estimated by the ejASMC algorithm at $r=0,1,9,18,\cdots,54$ iterations. The color of the posterior estimates changes from dark blue to light blue as more steps of SMC sampler are performed, decreasing the threshold $\varepsilon$. True values of parameters are indicated by vertical lines. }
\label{fig:GM1_SMC}
\end{figure}

In this study, we also
employ the ejASMC algorithm to infer the posterior distribution of the parameters, with $N=2048$ samples generated for each iteration and the threshold parameters are updated by retaining 1024 (the scale parameter $\gamma=0.5$) active particles. The termination condition for the algorithm is set such that the number of simulated data exceeds 50,000. As shown in Figure \ref{fig:GM1_SMC}, the posterior distribution becomes increasingly concentrated near the true value as the number of iterations increases.  {The threshold value $\varepsilon$ of the final iteration is $3.84$. ABC-MCMC algorithms achieve very low acceptance probability using this threshold. Here, the early rejection efficiency of ejASMC algorithm is 0.3820. $L_1$ distances between the marginal distributions of parameters and exact Bayesian posterior are 1.0626 and 1.0247, which are both smaller than the results provided by ejMCMC and OejMCMC (Appendix Table C.5), due to a smaller threshold. 
The posterior density plot provided by ejASMC are shown in Appendix Figure C.3.}

\subsection{Stochastic kinetic networks}

Consider the parameter inference of a multidimensional stochastic differential equation (SDE) model for biochemical networks \citep{golightly2010learning}. 
In the dynamic system, there are eight reactions (${R_1,R_2,\cdots,R_8}$) that represent a simplified model of the auto-regulation mechanism in prokaryotes, based on protein dimerization that inhibits its own transcription.
The reactions are defined as follows
$$\begin{aligned}
    R_1:&~ \text{DNA} + P_2\rightarrow \text{DNA} \cdot P_2 &
    R_2:&~ \text{DNA}\cdot P2\rightarrow \text{DNA} + P_2\\
    R_3:&~ \text{DNA}\rightarrow \text{DNA} + \text{RNA} &
    R_4:&~ \text{RNA}\rightarrow \text{RNA} + P \\
    R_5:&~ 2P \rightarrow P_2 &
    R_6:&~ P_2\rightarrow 2P\\
    R_7:&~ \text{RNA} \rightarrow \emptyset&
    R_8:&~ P \rightarrow \emptyset.  
\end{aligned}$$

Consider the time evolution of the system as a Markov process with state $\boldsymbol{X}_t$, where $\boldsymbol{X}_t=(\text{RNA},P,P_2,\text{DNA})^T$ represents the number of molecules of each species at time $t$ (\emph{i.e.} non-negative integers), and the corresponding stoichiometric matrix can be represented as
$$S=\left(
\begin{array}{cccccccc}
    0& 0& 1& 0& 0& 0& -1& 0\\
    0& 0& 0& 1& -2& 2& 0& -1\\
    -1& 1& 0& 0& 1& -1& 0& 0\\
    -1& 1& 0& 0& 0& 0& 0& 0
\end{array}
\right).$$
We refer readers to Section 6.2 of \cite{Picchini2014SDE} for a more detailed description of this networks.

Now we assume that a constant rate $\theta_i$ $(i=1,2,\cdots,8)$ is associated to each reaction $i$, the stochastic differential equation (SDE) for this model can be written as 
\begin{equation} d\boldsymbol{X}_t=\boldsymbol{S}\cdot\boldsymbol{h}(\boldsymbol{X}_t,\boldsymbol{\theta})dt+\boldsymbol{S}\sqrt{\text{diag}(\boldsymbol{h}(\boldsymbol{X}_t,\boldsymbol{\theta}))}d\boldsymbol{W}_t,
\end{equation}
where $\boldsymbol{h}(\boldsymbol{X}_t,\boldsymbol{\theta})=(\theta_1\cdot \text{DNA}\cdot P_2,\theta_2\cdot(k-\text{DNA}),\theta_3\cdot \text{DNA},\theta_4\cdot \text{RNA},\theta_5\cdot P\cdot(P-1)/2,\theta_6\cdot P_2,\theta_7\cdot \text{RNA},\theta_8\cdot P)^T$ is a ``propensity function", 
$k$ is a constant that satisfies $\text{DNA}\cdot P_2 + \text{DNA} = k$, 
and $h_i(\operatorname{X}_t,\theta_i)dt$ is the probability of reaction $i$ occurring in the time interval $(t,t+dt]$, $dW_t = (dW_{t,1},\ldots, dW_{t,8})^T$ is a Wiener process, where $dW_{t,i}$ is independently and identically distributed as $dW_{t,i}\sim N(0, dt)$ for $i = 1,\ldots, 8$. The goal is to infer the reaction rates $\boldsymbol{\theta}$ in different experimental settings. 


We consider three scenarios as in \cite{Picchini2014SDE}. $\mathcal{D}_1$: fully observed data without measurement error. $\mathcal{D}_2$: fully observed data with measurement error $\epsilon_{ij}\sim N(0,5)$ independently for every $i$ and for each coordinate $j$ of the state vector $\boldsymbol{X}_t$. $\mathcal{D}_3$: partially observed data (\emph{i.e.} the $\text{DNA}$ coordinate is unobserved) with measurement error $\epsilon_{ij}\sim N(0,\sigma^2 I_3)$, and $\sigma$ is unknown.


We use ejMCMC and OejMCMC algorithms to estimate the ABC posterior. The ABC-ASMC algorithm with OejMCMC move is used to collect $2000$ training data, with 500 particles, and $\varepsilon$ is updated by keeping 250 unique particles. In ABC-SMC, RMSE is used to model the discrepancy $\Delta$ between the observed data and the simulated data.

For $\mathcal{D}_1$ and $\mathcal{D}_2$, the threshold is set to $\varepsilon=0.03$. We use the Gaussian kernel function as the proposal distribution of MCMC move, and the covariance matrix of parameters with discrepancy less than 0.05 in the training data serves as the covariance parameter of the Gaussian kernel. 
For $\mathcal{D}_3$, we set $\varepsilon=0.05$, and training data with a discrepancy less than $0.06$ is used to tune the covariance matrix of the Gaussian kernel.

For all three scenarios, we run both ejMCMC and OejMCMC algorithms $500,000$ iterations. Table \ref{tab:SDE} shows the computing details of both algorithms. 
The results indicate that the efficiency of ejMCMC is higher than OejMCMC in all three scenarios.  In scenario $\mathcal{D}_2$, the ejMCMC algorithm can accelerate about 41.57\% compared with the OejMCMC algorithm.

\begin{table}[h]
    \centering
    \caption{Number of iterations ($N_{\text{ite}}$), number of predictions ($N_{\text{pre}}$), number of synthetic data generated ($N_{\text{sim}}$), number of accepted posterior samples ($N_{\text{acc}}$), early rejection efficiency for different scenarios.}
    \begin{tabular}{ccccccc}
    \hline
       & method&$N_{\text{ite}}$ &$N_{\text{pre}}$&$N_{\text{sim}}$&$N_{\text{acc}}$ &Eff \\ \hline
      \multirow{2}*{$\mathcal{D}_1$}
      & OejMCMC&500,000&-&124054&10574&0.7681\\
      &ejMCMC&500,000&123418&91580&10568 &\textbf{0.8345}\\ \hdashline
      \multirow{2}*{$\mathcal{D}_2$}
      & OejMCMC&500,000&-&120423 &7566&0.7708\\
      &ejMCMC&500,000&120367&73632&7935&\textbf{0.8665}\\ \hdashline
      \multirow{2}*{$\mathcal{D}_3$}
      & OejMCMC&500,000&-&299656&23185&0.4202\\
      &ejMCMC&500,000&298229&271459&20045&\textbf{0.4762}\\ \hline
    \end{tabular}
    \label{tab:SDE}
\end{table}

 {For three scenarios, the fitted GP discrepancy model is shown in Appendix Figure C.4-C.6, and the boxplots of the posterior samples are shown in Figure C.7-C.9 of Appendix C.2. The GP-ABC posterior (estimated using the training points) is less accurate than OejMCMC and ejMCMC since the GP model does not fit the discrepancy very well. Under scenario $D_3$, the true values of the parameters $\theta_3$ and $\theta_7$ are not within the 95\% credible interval of the OejMCMC algorithm, but are within those of the ejMCMC algorithm. The estimation results obtained by both algorithms in another two scenarios are similar, and the true values of the parameters are included in the estimated 95\% credible intervals.  We refer readers to Appendix C.2 for more details of setups and results.}

\section{Real Data Analysis}
\label{sec:real}

In this section, we consider a delay differential equation(DDE) example about the resource competition in the laboratory populations of Australian sheep blowflies ({\it Lucilia cuprina}), which is a classic experiment 
studied by \cite{nicholson1954outline}. The room temperature of cultivating blowflies were maintained at 25\textdegree{}C. 
 {Appendix Figure C.10 displays the counts of blowflies, which are  observed at $N = 137$  time points.} The resource limitation in the dynamic system of the blowfly population acts with a time delay, which causes the fluctuations shown in the blowfly population. The fluctuations displayed in the blowfly population are caused by the time lag between stimulus and reaction \citep{berezansky2010nicholson}. 

\cite{may1976models} proposed to model the counts of blowflies with the following DDE model
\begin{eqnarray}
\frac{dx(t)}{dt} & = & \nu x(t)[1 - x(t-\tau)/(1000\cdot P)],
\end{eqnarray}
where $x(t)$ is the blowfly population, $\nu$ is the rate of increase of the blowfly population, $P$ is a resource limitation parameter set by the supply of food, and $\tau$ is the time delay, roughly equal to the time for an egg to grow up to a pupa. Our goal is to estimate the initial value, $x(0)$, and the three parameters, $\nu$, $P$, and $\tau$, from the noisy blowfly data $y(t)$. The observed counts of blowflies $y(t)$ is assumed to be lognormal distributed with the mean $x(t)$ and the variance $\sigma^2$. 

For convenience of description, we assume $\thetabold=\{X_0,v,P,\tau\}$, and let $\ybold$ denote the observed data.  We define the following prior distributions for $\thetabold:~\log(X_0)\sim\mathcal{N}(8.5,0.3^2),~\log(v)\sim\mathcal{N}(-1.35,0.2^2),~\log(P)\sim\mathcal{N}(0.8,0.3^2),~\log(\tau)\sim\mathcal{N}(2.25,0.08^2)$. We use root mean square error as the discrepancy function $\Delta$ for ABC.
The simulation data are
generated via the Euler-Maruyama method \citep{maruyama1955E-M} implemented in R package \emph{deSolve} \citep{soetaert2010solving}, using a constant step-size 0.1.  {For the cheap simulator of DAMCMC algorithm, we set the step-size to 2 for the first forty time units, and set the step size of the rest time unit to the time interval of the observation data.}

We collect $3000$ training data by 
running an ABC-ASMC algorithm  with $500$ particles, and use the resulting parameter-discrepancy pairs to train the GP discrepancy model.
The predict function $h_a(\thetabold)$ is the GP model trained using the $3000$ samples with $a=0.05$. 
 {We run {ejMCMC}, {OejMCMC} and DAMCMC $500,000$ iterations.} 
The {ejMCMC} algorithm  simulates $110,564$ synthetic data and saves $93.09\%$ of the cost, while {OejMCMC} algorithm simulates $291,250$ synthetic data and saves only $49.91\%$ of the cost. Our algorithm {ejMCMC} significantly improve the computational speed compared with OejMCMC. Both algorithms accept approximately $82,000$ samples.   {The DAMCMC algorithm runs the cheap simulator $500,000$ times and the expensive simulator $82,692$ times, and accepts $66,028$ samples. In addition, we compare the computing time of three MCMC methods as a function of the step size. The results are reported in Appendix Section C.3.}




\begin{figure}[h]
    \centering
    \includegraphics[width=0.8\linewidth]{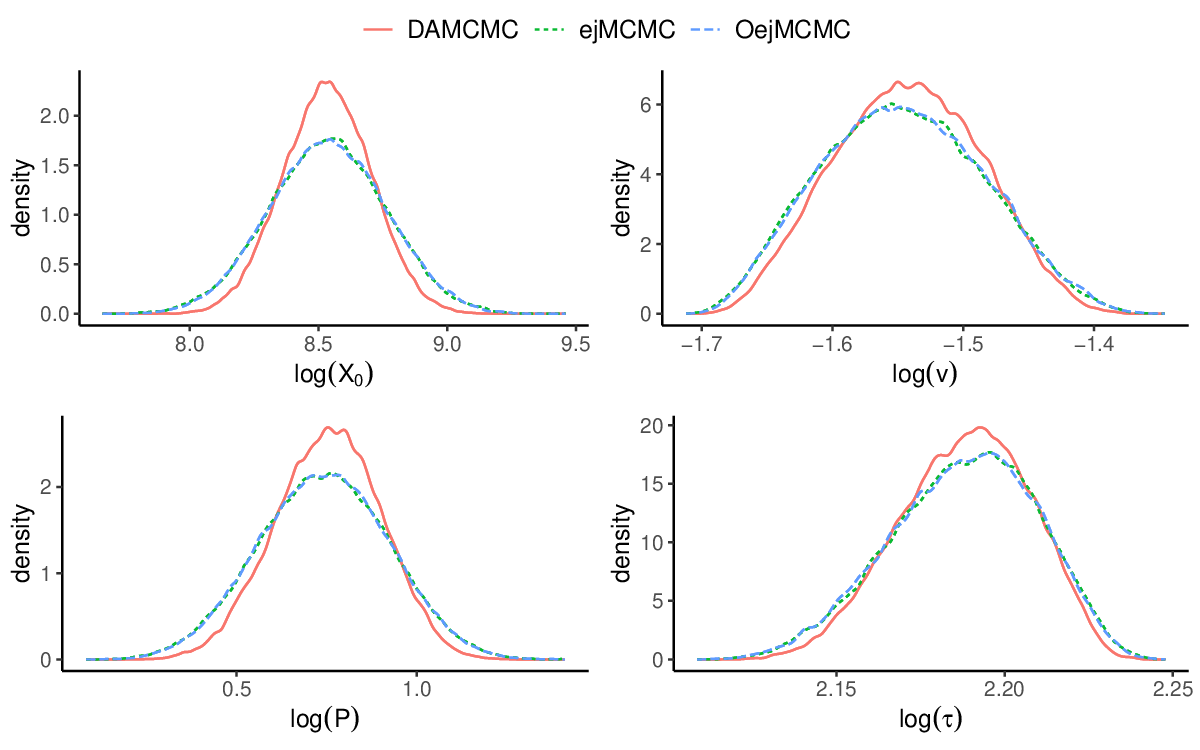}
    \caption{ {Marginal ABC posterior distributions estimated by OejMCMC, ejMCMC and DAMCMC.}}
    \label{fig:realdata}
\end{figure}
\begin{figure}[ht]
    \centering
    \includegraphics[width=1\linewidth]{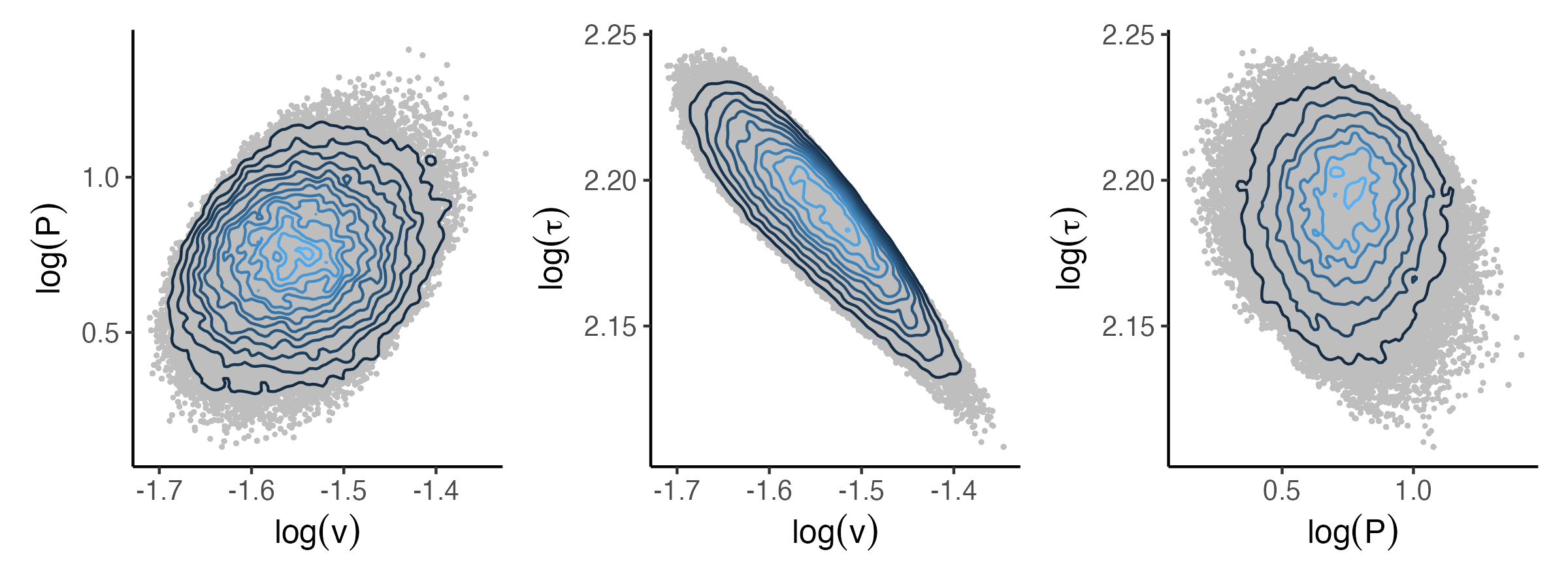}
    \caption{The posterior distributions of parameters estimated by {ejMCMC} with $500,000$ MCMC iterations. Grey points: the accepted particles; loop lines: the 2D joint density contours. }
    \label{fig:Real data density 2d ejMCMC}
\end{figure}

 {Figure \ref{fig:realdata} displays the marginal ABC posterior distributions produced by the OejMCMC, ejMCMC and DAMCMC algorithms. Figure \ref{fig:realdata} shows that the posterior of ejMCMC algorithm is consistent with that of  OejMCMC algorithm. However, the posterior generated by the DAMCMC algorithm differs from that of OejMCMC and ejMCMC algorithms.} Figure \ref{fig:Real data density 2d ejMCMC} shows the two-dimensional joint posterior distributions of the parameters obtained by using the {ejMCMC} algorithm.  {The posterior means and 95\% credible intervals of $\thetabold$ provided by the OejMCMC, ejMCMC and DAMCMC algorithms are shown in Appendix Figure C.6.}
We also compute the correlation between posterior samples: $corr(\nu, P) = 0.264$, $corr(\nu, \tau) = -0.882$, $corr(P,\tau) = -0.112$. 
Recall that $\nu$ is the rate of increase of the blowfly population, $P$ is a resource limitation parameter set by the supply of food, and $\tau$ is the time delay that roughly equal to the time for an egg to grow up to a pupa. The positive correlation between $\nu$ and $P$ indicates that the blowfly population grows faster when there is a sufficient food supply. The tiny negative value of the correlation between $\tau$ and $P$ implies that the amount of food supply has a small impact on the period of being a pupa.
The negative correlation between $\tau$ and $\nu$ indicates that the blowfly population will increase slower if the period for an egg to grow up to a pupa is longer.

\section{Conclusion}
\label{sec:conc}
We propose the early rejection ABC methods to accelerate parameter inference for complex statistical models. The main idea is to accurately reject parameter samples before generating synthetic data, and this is achieved by using a Gaussian process modeling to describe the relationship between the parameter and discrepancy. Our proposed method has two advantages compared with existing work. First, it utilizes MCMC moves for sampling parameters, which is more  efficient in high-dimensional cases than sampling from a prior distribution. Second, it uses a pseudo Metropolis Hasting acceptance ratio to early reject samples in state space that are not worth exploring. The new method inherits the detailed balance condition and the posterior concentration property. It is also theoretically more efficient than the existing early rejection method. 

We use several schemes to improve the performance of {ejMCMC}. To ensure the accuracy of the discrepancy model, we sequentially generate training points in areas with smaller discrepancy via a pilot run of ABC-SMC. We propose a prediction function for discrepancy model that balances accuracy and computational speed. An adaptive early rejection SMC is introduced by combining ABC-SMC and {ejMCMC} method. The ejASMC can adaptively select the sequence of thresholds and MCMC kernels, the particles are initialized via uniform design. 

 {There are several lines of extensions and improvements for future work. 
We utilize identical kernel function and threshold value for $\Delta(x, y)$ and $h(\thetabold)$ in the pseudo acceptance ratio ({\it i.e.} $\min \left\{K_\varepsilon(\Delta(x, y)), K_\varepsilon(h(\thetabold))\right\}$) of ejMCMC. Using different kernel functions and threshold values can enhance the flexibility of algorithm. For example, decreasing the threshold of $K_\varepsilon(h(\thetabold))$ can decrease the rate of false positive. One future extension is to propose pseudo acceptance ratio with more general kernel functions. MCMC approaches often require a large number of simulations to make sure that the Markov chain converges to the stationary distribution. This limits their application to models that only a few hundreds simulations are possible.
Compare to MCMC methods, importance sampling does not require to assess convergence and the consistency property holds under mild conditions. The efficiency of importance sampling methods mainly depends on proposal distributions. Our second line of future research is to combine GP-ABC and importance sampling approaches, by serving the fitted GP model as an efficient proposal distribution in importance sampling framework. In addition, we will investigate using the resulting importance sampling distribution to design efficient MCMC proposal distributions to improve the mixing of Markov chain.
The current selection of discrepancy model is limited to models defined on Euclidean space. Lastly, we will extend $h(\thetabold)$ to problems defined on more general space (\eg a phylogenetic tree) in our future research. 
} 

~~
\section*{Acknowledge}
This work was supported by the National Natural Science Foundation of China (12131001 and 12101333), the startup fund of ShanghaiTech University, the Fundamental Research Funds for the Central Universities, LPMC, and KLMDASR. The authorship is listed in alphabetic order.

\bigskip
\begin{center}
{\large\bf SUPPLEMENTAL MATERIALS}
\end{center}

\begin{description}

\item[Appendix A:] The proofs of all theoretical results presented in the main text.

\item[Appendix B:]  {Some description of algorithms not shown in the main text.}

\item[Appendix C:]  {Some numerical results and some details of setups not shown in the main text.}

\end{description}

\bibliographystyle{chicago}
\bibliography{Bibliography-MM-MC}
~~

\end{document}